\documentclass[a4paper,12pt]{article}
\usepackage[utf8]{inputenc}
\usepackage[english]{babel}
\usepackage{authblk}
\usepackage{graphicx}
\usepackage{mathptmx}
\usepackage[singlespacing]{setspace}
\usepackage[headheight=1in,margin=1in]{geometry}
\usepackage{fancyhdr}
\usepackage{lipsum}
\usepackage{apacite}
\usepackage{longtable}
\usepackage{tabularx}
\usepackage{pgfplots}
\usepackage{color,soul}
\usepgfplotslibrary{groupplots}
\usepackage{tikz}
\usetikzlibrary{arrows,backgrounds,snakes}
\pgfplotsset{compat=1.18}
\usepackage[hyphens]{url}

\usepackage{CJKutf8}

\usepackage{todonotes}

\usepackage{todonotes}

\usepackage{array}

\makeatletter
\def\@maketitle{%
  \newpage
  \begin{center}%
    \let \footnote \thanks
    {\LARGE \@title \par}%
    \vskip 1em
    {\large \@author \par}
  \end{center}%
  \par
  \vskip 0.1em
  }
\makeatother

\graphicspath{{images/}}

\title{Predicting Depressive Symptoms through Emotion Pairs within Asian American Families}

\author[1]{Sangpil Youm}
\author[2]{Nari Yoo}
\author[3]{Sou Hyun Jang}
\affil[1]{University of Florida, Gainesville, FL, USA}
\affil[2]{University of Michigan, Ann Arbor, MI, USA}
\affil[3]{Korea University, Seoul, Korea}

\begin{document}

\maketitle
\thispagestyle{fancy}

\begin{center}
\newline
\end{center}

\section*{Extended Abstract}

Asian Americans represent the most rapidly expanding racial/ethnic group in the United States, with their population nearly doubling from 2000 to 2019 \cite{Budiman2021-zu}. The intergenerational relationship between parents and children in Asian American families is a complex and multifaceted dynamic and has received much attention from scholars \cite{Warikoo2020AddressingAmerica1}. The intergenerational relationship within families is characterized by mixed or ambiguous feelings, called \textit{intergenerational ambivalence} \cite{Weng2014-bi, Muruthi2017-dq}. Understanding the nature of mixed feelings is crucial, as it can inform how we approach and manage this complex emotional landscape. Previous studies have revealed that ambivalence within Asian American families presents a spectrum of experiences and that this ambivalence has an impact on both psychological well-being and physical health \cite{Hebblethwaite2015UnderstandingPackage}.

While earlier studies have uncovered ambivalence within Asian American families, preliminary investigations have been limited to a narrow range of people through qualitative methods, such as surveys and interviews. To address this limitation, the current study extracts emotions from Reddit posts to consider a richer context in the narrative and to explore the effects of emotional ambivalence on depressive symptoms. Furthermore, our study establishes a connection between emotion pairs and depressive symptoms, providing insight into the complexities of depressive symptoms within a broader context.

Reddit allows user to share more candid and potentially sensitive personal stories, which can be considered more authentic due to reduced fear of social repercussions \cite{Choudhury2014MentalAnonymity}. For our study, we collected 31,144 posts from subreddit \textit{r/AsianParentStories} spanning a decade, from 2012 to 2022. This subreddit is a specifically designed for user to share stories that often portray their Asian parents in a negative light, seeking empathy and understanding from a community of peers with similar background.

\begin{figure}[htp]
\centering
\includegraphics[width=10cm]{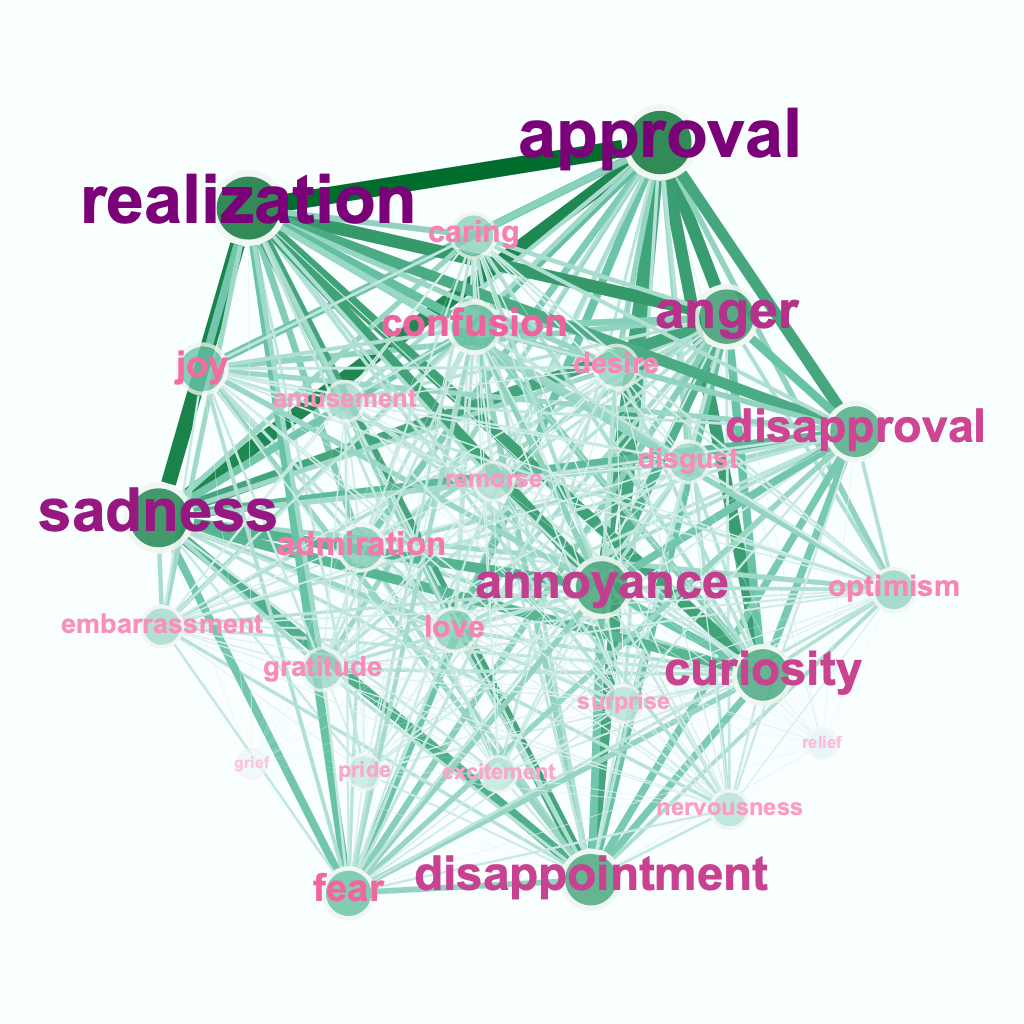}
\caption{Emotion relationship network. Note. Nodes represent emotions. Link thickness corresponds to size of degree, measured by frequency of co-occurrence}
\label{fig:network}
\end{figure}

To conduct a comprehensive analysis of posts, first, we employ emotion detection at the sentence level using EmoRoBERTa \cite{Kamath2022AnRoBERTa}. Subsequently, we gauge the co-occurrence of emotions in a post, forming an emotional network. In this network, each node represents an emotion presented in the post, and the degree of a node corresponds to the frequency of its occurrence in the post. Thicker links between nodes signify a higher frequency of co-occurrence between two emotions. This emotion network is visually depicted in Figure~\ref{fig:network}. 

Additionally, we explore the relationship between emotion pairs and depressive symptoms, as these emotion pairs serve as factors influencing depressive symptoms. To identify depressive symptoms within each post, we employ the DepRoBERTa \cite{Poswiata2022OPILT-EDI-ACL2022:Models}. We then treat the emotion pairs as variables and depressive symptoms as the predicted value. Through the application of logistic regression, we extract statistically significant emotion pairs. Next, we apply content analysis to gain further insights and interpretations of these emotion pairs.

\begin{figure}[htp]
\centering
\includegraphics[width=15cm]{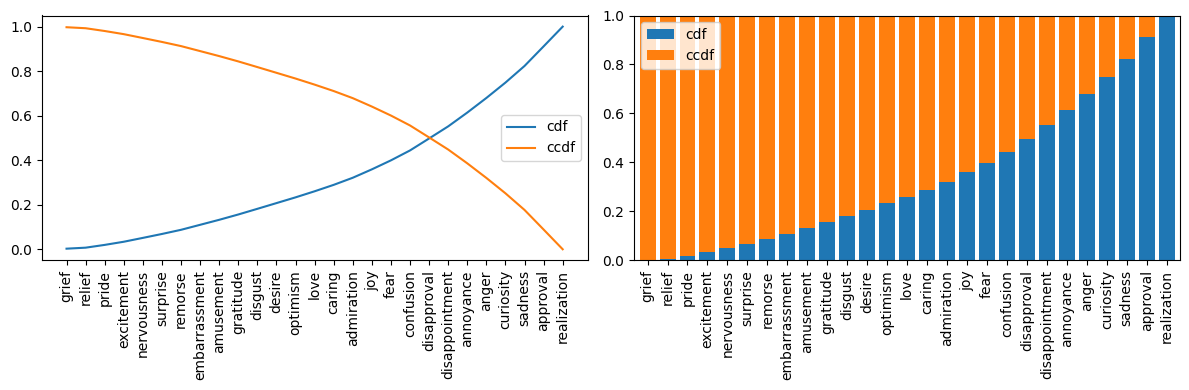}
\caption{Cumulative Distribution Function (CDF) and Complementary Cumulative Distribution Function (CCDF).}
\label{fig:cdfccdf}
\end{figure}

This study yields three significant implications. First, among the over 28 detectable emotions across all sentences in posts, eight emotions—namely, \textit{realization, approval, sadness, anger, disapproval, annoyance, curiosity,} and \textit{disappointment}—comprise 50\%. The Cumulative Distribution Function (CDF) and Complementary Cumulative Distribution Function (CCDF) in Figure~\ref{fig:cdfccdf} demonstrate the dominance of these eight emotions, collectively constituting half of the emotional content. Moreover, these 8 emotions exhibit a high degree of interactivity among themselves and with other emotions through network, as depicted in Figure~\ref{fig:network}. This outcome underscores the interconnectedness of emotions. The investigation into the relationships between emotions holds significant implications for the effective analysis of posts.

\begin{figure}[htp]
\centering
\includegraphics[width=12cm]{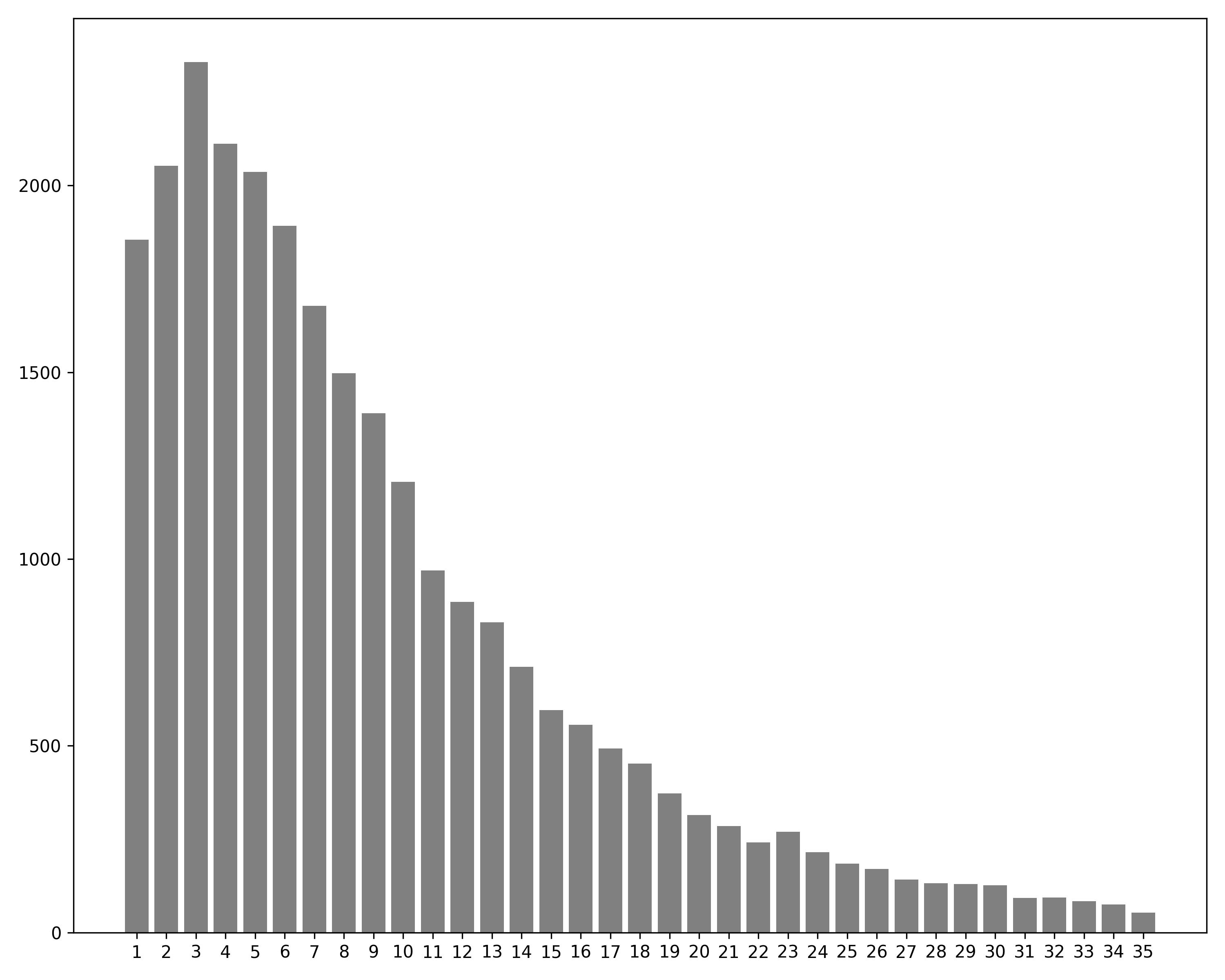}
\caption{Distribution of the number of pairs per post. Note. The X-axis represents the number of pairs, and the Y-axis represents the frequency of pairs}
\label{fig:distPairs}
\end{figure}

\begin{figure}[htp]
\centering
\includegraphics[width=15cm]{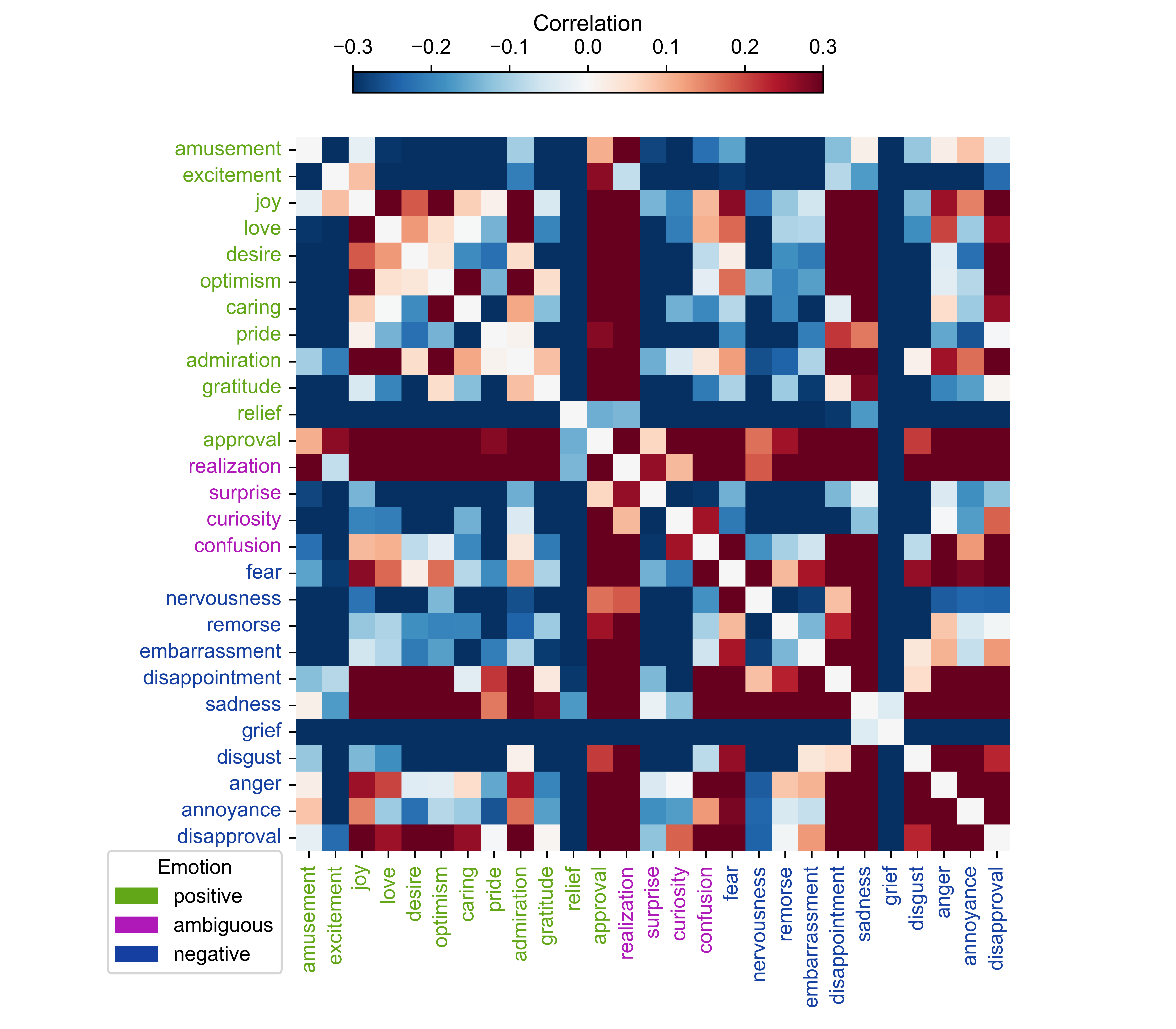}
\caption{Co-occurrence Matrix. Note. Blue-colored emotions denote negative emotions and green-colored emotions denote positive emotions and purple-colored emotion denote neutral emotion}
\label{fig:heatmap}
\end{figure}

Second, the ambivalence of emotions becomes more apparent when analyzing the co-occurrence of emotions. As illustrated in Figure~\ref{fig:distPairs}, it is evident that more than two emotions are detected per post, with pairs and triplets of emotion combinations being the most frequently observed. To comprehensively analyze ambivalent emotions in a post, we consider all permutations of emotions, treating pairs as the fundamental unit for emotion-depressive symptoms analysis. These co-occurrences of emotions are not limited to those of the same sentiment; they also extend to different sentiment emotions, as demonstrated in the heatmap in Figure~\ref{fig:heatmap}.

Third, our analysis, leveraging emotion pairs, uncovers that the complex emotion pair \textit{amusement-grief} exerts the most significant positive effect on depressive symptoms. Conversely, the emotion pair \textit{caring-curiosity} exhibits a most negative impact on depressive symptoms. We identify 10 emotion pairs that significantly contribute to depressive symptoms in Table.~\ref{tab:coef_exp}, falling within a significance level of 0.05\%. The odds ratio for each emotion pair indicates whether the respective emotions positively or negatively influence depressive symptoms.

Our findings regarding the relationship between emotion pairs and depressive symptoms reveal that the impact of an emotion depends on its co-occurrence with other emotions. For instance, while \textit{amusement} is generally a positive emotion that might have a negative effect on depressive symptoms, when it co-occurs with \textit{grief} in a post, this pair has a positive effect on depressive symptoms, with a high odds ratio of 2.62 as depicted in Table.~\ref{tab:coef_exp}. Interestingly, \textit{amusement} in the context of Asian parents are more close to sarcasm, not a dictionary of \textit{amusement} meaning of enjoyment. For instance, \textit{\textbf{[grief]} I have found myself seriously considering ending my life on several occasions ...
The pressure exerted by parents is a real thing, yet it typically does not generate feelings of hatred ... \textbf{[amusement]} 
Parents and children playfully joke about past moments of discipline, sharing laughter over those experiences. }

Similarly, \textit{sadness}, typically a negative emotion that might positively affect depressive symptoms, shows a negative effect on depressive symptoms when paired with \textit{optimism}, with an odds ratio of 0.79 as depicted in Table.~\ref{tab:coef_exp}. \textit{sadness} represent the empathy of their parents. For instance, \textit{\textbf{[sadness]} My moms life has been exhausted - A single parent of two, divorced, experienced continental migration, cultural shocks, relatives' deaths, self-abnegation, miscarriage, and countless sacrifices ... She doesn't understand my career path ... \textbf{[optimism]} We are not same person, and I simply want to explore my own career path}.

Our findings demonstrate that mixed emotions uncover a more intricate context within Reddit posts related to Asian parents in explaining depressive symptoms among Asian children rather than assuming single and consistent emotions among people. This study has practical and clinical implications for understanding Asian Americans in the context of parent-child relationships and guiding culturally informed family therapy \cite {Natrajan-Tyagi2018-js}. We suggest that future studies delve into intergenerational ambivalence among other racial/ethnic communities, such as Latinos \cite {Pei2019-no}, who represent another significant immigrant population. It would be beneficial to compare the cultural and contextual factors linked to mixed emotions and mental health across these groups, aiming to develop effective interventions.

\newpage
\Urlmuskip=0mu plus 1mu
\bibliographystyle{apacite}
\bibliography{references, custom.bib}

\newpage
\section*{Figures and Tables}

\begin{table}[htp]
\centering
\includegraphics[width=11cm]{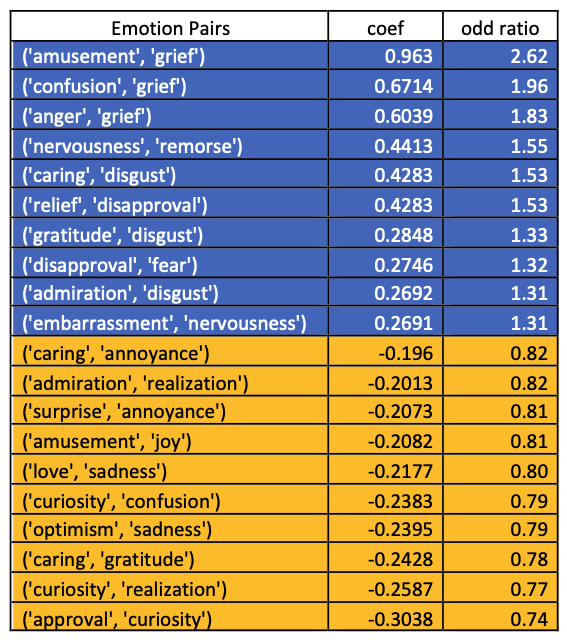}
\caption{Emotion pairs significance with coefficient and odd ratio. Note. An odd ratio with over 1 indicates positive impact on predicted value (\textit{depressive symptoms)}, while an odd ration between 0 - 1 indicates negative impact on predicted value (\textit{depressive symptoms)}}
\label{tab:coef_exp}
\end{table}


\end{document}